# Deep Reinforcement Learning Based Toolpath Generation for Thermal Uniformity in Laser Powder Bed Fusion Process

Mian Qin [a], Junhao Ding [a], Shuo Qu [a], Xu Song [a, b], Charlie C.L. Wang [c], Wei-Hsin Liao [a, b]

[a] Department of Mechanical and Automation Engineering, The Chinese University of Hong Kong, Shatin, Hong Kong, China

[b] Institute of Intelligent Design and Manufacturing, The Chinese University of Hong Kong, Shatin, Hong Kong, China

[c] Department of Mechanical, Aerospace and Civil Engineering, The University of Manchester, Manchester, UK

## Abstract

Laser powder bed fusion (LPBF) is a widely used metal additive manufacturing technology. However, the accumulation of internal residual stress during printing can cause significant distortion and potential failure. Although various scan patterns have been studied to reduce possible accumulated stress, such as zigzag scanning vectors with changing directions or a chessboard-based scan pattern with divided small islands, most conventional scan patterns cannot significantly reduce residual stress. The proposed adaptive toolpath generation (ATG) algorithms, aiming to minimize the thermal gradients, may result in extremely accumulated temperature fields in some cases. To address these issues, we developed a deep reinforcement learning (DRL)-based toolpath generation framework, with the goal of achieving uniformly distributed heat and avoiding extremely thermal accumulation regions during the LPBF process. We first developed an overall pipeline for the DRL-based toolpath generation framework, which includes uniformly sampling, agent moving and environment observation, action selection, moving constraints, rewards calculation, and the training process. To accelerate the training process, we simplified the data-intensive numerical model by considering the turning angles on the toolpath. We designed the action spaces with three options, including the minimum temperature value, the smoothest path, and the second smoothest path. The reward function was designed to minimize energy density to ensure the temperature field remains relatively stable. To verify the effectiveness of the proposed DRL-based toolpath generation framework, we performed numerical simulations of polygon shape printing domains. In addition, four groups of thin plate samples with different scan patterns were compared using the LPBF process. The results indicated that using our proposed DRL algorithms for toolpath generation reduced the maximum distortion of the sample by approximate 47% compared to Zigzag patterns, 29% compared to Chessboard patterns, and 17% compared to ATG patterns. Our work demonstrated the effectiveness of implementing machine learning algorithms in the toolpath generation of the LPBF process.

## Keywords

Laser powder bed fusion (LPBF); Reinforcement learning; Toolpath generation; Numerical simulation; Scan pattern

## 1. Introduction

Laser powder bed fusion (LPBF) is a commonly used metal additive manufacturing (AM) technology that fabricates parts on a powder bed using high-energy laser scanning layer-by-layer, producing highly complex structures and components with excellent mechanical properties [1]-[4]. However, the accumulation of internal residual stress during printing may cause large distortion and potential failure. The choice of scan pattern can significantly affect the local thermal condition and thermal stress [5]-[8].

While zigzag scan patterns have been commonly used due to their stability and simplicity, they can significantly increase thermal stress and lead to excessive distortion [9]-[11]. Previous studies by Yang et al. [12] and Catchpole-Smith et al. [13] have explored the fractal scanning pattern, but they lacked theoretical guidance and practical implementation. Recently, as the thermal gradient has been identified as the primary cause of residual stress and distortion [14], several toolpath optimization approaches based on thermal-mechanical models have been proposed. Boissier et al. [15] explored temperature field changes in scan path optimization but encountered variable hatch space, potentially leading to non-uniform or incomplete melting. Chen et al. [16] introduced an island-based scan patterns generation method using finite element analysis (FEA) with approximated voxels, which demands significant computing resources, especially when dealing with varying cross-sections. Ramani et al. [17] proposed the "smartscan" method for improving thermal uniformity in zigzag and island patterns, but their algorithms were only tested on simple regular boundaries. Chen et al. [18] introduced a continuous laser scanning optimization method aimed at reducing residual stress for a given geometry, but their studies were only implemented on simulated models. Takezawa et al. [19] explored the optimization of hatching orientation and lattice density distribution, considering layer-wise residual stress stacking. Their method was validated through both numerical simulations and experiments. Nevertheless, traditional zigzag scan patterns were adopted in each layer, and thermal uniformity was not optimized. Qin et al. [20] proposed the adaptive toolpath generation (ATG) algorithms, which minimizes the thermal gradients by finding the "best" next point within the searching regions under fabrication constraints. However, this algorithm may result in extreme accumulated temperature fields in some cases.

Deep reinforcement learning (DRL) algorithms, as a type of machine learning method, can obtain the optimal policy by maximizing the accumulated rewards through interaction with the environment [21]-[26]. Recent studies have explored the application of DRL algorithms in the field of metal additive manufacturing. Dharmawan et al. [27] proposed a reinforcement learning-based control strategy to improve the surface quality of parts fabricated by their robotic metal wire arc AM platform. In their work, they treated the points on the toolpath as agents that move along the zigzag scan patterns. However, their studies did not focus on optimizing the thermal distribution of toolpath, particularly for complex toolpaths with adaptive features. Ogoke and Farimani [28] explored the use of DRL algorithms as an effective control policy for thermal control in the LPBF process. Their research mainly centered around optimizing the parameters such as laser power and velocity during the fabrication process.

To address the above-mentioned issues, we propose a DRL-based scan pattern generation framework, aiming to achieve uniformly distributed temperature fields while avoiding extreme cases with accumulated thermal regions. Within this framework, the laser beam acts as a mobile agent, and the agent's actions correspond to the movement of the laser beam in the LPBF process. By employing DRL algorithms to control the agent, we can formulate the scan pattern generation process as an optimization problem, seeking the most uniform distribution of temperature fields within the printing domains. The elimination of extremely accumulated thermal regions will be accomplished by maximizing the reward function, which can be designed to minimize the energy density as much as possible.

In this paper, our proposed approach follows an overall pipeline that includes uniform sampling, agent movement, environment observation, action selection, moving constraints, reward calculation, and training. The DRL-based algorithm focuses on minimizing energy density to ensure a stable temperature field and avoid extreme thermal accumulation. The action spaces are designed to offer three options: minimum temperature value, smoothest path, and second smoothest path. We conducted numerical simulations to compare with other scan patterns in terms of average and maximum depth values of the molten pool. Additionally, we conducted experiments by implementing four groups of thin plate samples with different scan patterns, including Zigzag patterns, Chessboard patterns, ATG patterns and DRL patterns. Overall, our DRL-based approach provides a promising solution to address the problem of residual stress accumulation during the LPBF process. This work highlights the potential of machine learning algorithms in optimizing toolpaths for additive manufacturing processes.

## 2. Algorithms design

### 2.1 overall pipeline

The overall working pipeline of the DRL-based scan pattern generation framework is presented in Fig. 1. The training process is utilized to obtain the optimal policy (Fig. 1 (a)), and then the optimal toolpath pattern is generated based on the optimal policy, as shown in Fig. 1 (b). We adopt the deep Q-network (DQN), which combines reinforcement learning and artificial neural network, as reported by Wolfer et al. [29]. Neural networks are trained during the iteration of scan pattern generation over a set number of episodes. The pseudocode of our proposed algorithms is provided in Algorithm 1. The overall pipeline mainly consists of the following steps:

***1) Uniformly sampling:*** In our research, the input printing domain is filled with a set of uniformly distributed sample points. These sample points enable the optimized scan patterns to be obtained by traversing all points within the printing domain. The distance between adjacent points, also known as the hatch space $h$, is set to 50 $\mu m$, which is determined by the fabrication machine. As a result, the distribution of sampled points is related to the boundary and size of the printing domain. The initial sampled points are shown on the left side of Fig. 1 (b) with the distance between adjacent points set to the hatch space.

***2) Agent moving and environment observation:*** Our DRL-based scan pattern generation framework employs a moving laser beam as the mobile agent, which is the main object of investigation during the training process. Real-time numerical simulations are integrated into the system to carry out the training process offline, and as a result, the agent (laser beam) is simulated instead of using a real one.

We use the temperature fields of the powder bed, calculated by data-intensive numerical simulation methods, as the environment. The agent can move step by step during training through action section operation. Once the agent successfully moves to a new coordinate, the corresponding environment information is immediately updated using numerical simulation results. To simplify the numerical environment, we minimize the energy density and investigated the relationship between the turning angles and the maximum depth values, as shown in Fig. 2. We will introduce data-intensive numerical simulation in Section 2.2.

***3) Action selection and moving constraints:*** To connect all initially sampled points, the movement of the agent needs to be carefully defined in the action selection strategy. The agent's movement is constrained by the boundary of the printing domain, including the free agent, boundary agent, and corner agent. Three typical environments of agents with different possible actions are illustrated in Fig. 3 (a) - (c). Then, based on these environments, we design the action spaces with three selection strategies: the minimum temperature value, the smoothest path, and the second smoothest path, as shown in Fig. 4.

To ensure the quality of the generated toolpath patterns, we define two constraints: "unqualified-point" and "isolated-point," as explained in Section 2.6. In our proposed algorithm, we impose extra punishment in the reward function calculation for collision more than a certain number of times or isolated-points.

***4) Reward calculation:*** Once the agent successfully takes an action, the updated environment is observed, and reward values can be calculated based on the real-time states of the agent. Efficient numerical simulation methods are used to calculate the new environment states. The actions, states, and rewards are then stored and used for training in the next step. The detailed reward functions will be introduced in Section 2.4.

***5) Training:*** Once the rewards are calculated, which are directly related to new actions and states, DQN model will be updated. This process will be performed iteratively until the candidate points set is empty. During the training process, if all the points have been traversed, the current training episode will be ended immediately. The total number of episodes is determined based on experimental practice, which is related to both the parameters of the algorithms and the shape of printing domains. The training process in this research will be discussed in detail in Section 3.1.

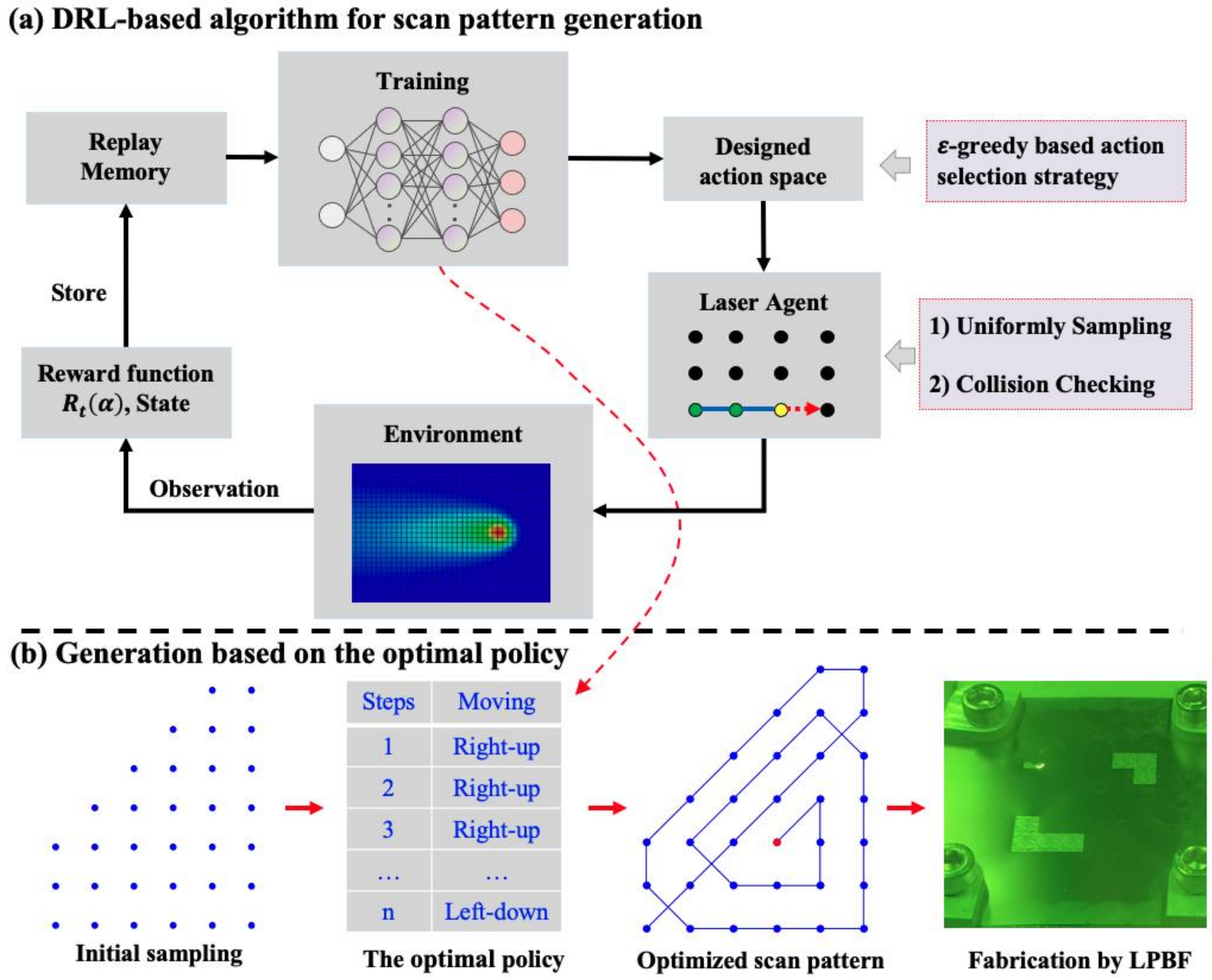

Fig. 1 Overall pipeline of DRL-based algorithm for toolpath optimization (a) Algorithms and training process of DRL-based algorithm (b) Generation of optimized scan pattern using the optimal policy.

---

**Algorithm 1: DRL-based algorithm for scan pattern generation**

---

**Input:** Printing domain with dimensions at $(x, y)$, hatch distance $h$, collision threshold $N$, action space $A(s)$
**Output:** Optimized toolpath graph $\bar{\mathcal{G}}$

Uniformly sample points in Environment $S(x, y)$ and initialize the agent
Initialize main network $Q(s, a)$ and target network $Q_t(s, a)$ with s ∈ $S$, a ∈ $A(s)$
Initialize experience replay memory $M$
***for*** each episode ***do***
    Initialize $S_t$
    ***for*** each step of current episode ***do***
        Choose action $A_t$ from $S_t$ using ε-greedy strategy
        Check collision of action $A_t$
        ***if*** collision: count collision times $N_{ct}$
            ***if*** $n > N$ *or* is isolated-point: $R_t \leftarrow R_t - 1$
        Take the action $A_t$ and observe $S_{t+1}$
        Calculate the reward function $R_t(d)$
        Store transition $(S_t,\ A_t, R_t,\ S_{t+1})$ in memory $M$
        Randomly choose minibatch $B$ from $M$
        $Q(S_{t+1}, a) = \begin{cases} R_t & terminal \\ R_t + \gamma \max_a \left(\mathrm{Q}\ (S_{t+1}, a)\right) & non-terminal \end{cases}$
        Perform gradient descent algorithm to minimize the loss
        Update $Q_t(s, a) \leftarrow Q(s, a)$ with frequency *f;* $S_t \leftarrow S_{t+1}$
    ***Repeat until*** number of steps equals to number of sampled points
***Repeat until*** converged and store the optimal policy $\pi^*$
***Return*** the optimal toolpath graph $\bar{\mathcal{G}}$ based on $\pi^*$

---

## 2.2 Data-intensive numerical simulation

During the training of toolpath optimization algorithms, it is essential to utilize real-time temperature field distributions. Therefore, the numerical simulation process should be efficient and feasible, enabling frequent input and output of data. However, current modeling methods commonly employed in commercial applications often demand a substantial allocation of computational resources, which is unsuitable for data-intensive optimization of adaptive toolpaths.

Wolfer et al. [30] developed the RUSLS (Repeated Use of Stored Line Solutions) method to quickly predict solutions for the thermal history of the molten pool in the numerical model, which mainly calculates the transient heat conduction of moving laser beams on the top surface of the powder bed. Their numerical simulations can accommodate arbitrary scanning patterns, which is suitable for adaptive toolpath patterns. Ogoke and Farimani [28] have conducted experiments to verify the numerical model by comparing the simulated and experimental depths using both SS316L and Ti-6Al-4V materials. The key equations used in the RUSLS numerical model are shown below [28].

The equation of the heat flux on the top surface is given by:

$$\frac{\partial \mathrm{T}}{\partial \mathrm{n}} = \frac{\mathrm{k}}{\rho \mathrm{c_p}} \nabla^2 \mathrm{T}(\mathrm{x}, \mathrm{y}, \mathrm{z}, \mathrm{t}) - \frac{\mathrm{Q_{input}}}{\rho \mathrm{c_p}} \tag{1}$$

where $k$ is thermal conductivity; $\rho$ is the density of material; $\mathrm{c_p}$ is heat capacity of material; $\mathrm{Q_{input}}$ is the input heat source.

To describe temperature field $T$, the heat source term $T_Q^{(i)}$ and heat diffusion history term $T_D^{(i)}$ are defined, respectively:

$$T^{(i)}(x, y, z) = T_Q^{(i)}(x, y, z) + T_D^{(i)}(x, y, z) \tag{2}$$

For simplicity, laser heat source is regarded as Gaussian distribution:

$$\Theta(x, y, z, t) = \frac{AP}{2\pi\sigma^2 \rho c_p} \exp\left(-\frac{(x-vt)^2+y^2}{2\sigma^2}\right)\delta(z) \tag{3}$$

where $\Theta = \frac{Q_{input}}{\rho c_p}$; A is the absorptivity; P is the power; σ is the laser diameter; v is the scanning velocity.

However, the RUSLS numerical model is still computationally demanding. We perform numerical simulations at each step, utilizing the updated state (coordinate) of the agent (laser beam) as input within the DRL framework. This resource-intensive process is chiefly attributed to the substantial number of iterations required in our developed DRL framework. To further enhance computational efficiency, this study aims to investigate the correlation between turning angle and thermal accumulation in adaptive scan patterns, as depicted in Fig. 2. This relationship is crucial since the thermal distribution is highly sensitive to turning angle. The real-time turning angle $\alpha$ of adaptive scan patterns has been normalized between 0 and 180. To simulate the most common adaptive scan patterns, we have used the RUSLS approach to create eight typical template situations, which are shown in Fig. 2 (a), while the corresponding numerical results are presented in Fig. 2 (b). The toolpath visualization has been represented by blue and red lines, where blue lines show the previous toolpath, and red lines show the next possible toolpaths. The maximum depth of the molten pool has been used as a metric for comparing thermal accumulation.

Our findings, as shown in Fig. 2 (c), indicate that the relationship between turning angle and thermal accumulation in adaptive scan patterns is highly sensitive. When the turning angle $\alpha$ is less than 90 degrees, the depth values of the molten pool increase significantly as the turning angle decreases. However, when α is greater than 90 degrees, the depth values remain constant at around 45 $\mu m$. Therefore, we have proposed a sensitive region in adaptive scan patterns that occurs when two successive turning angles are less than 90 degrees, and the length between them is smaller than a particular value, as defined in ***Definition 1***. The coefficient of the sensitive region determines this value, which is related to the fabrication environment of the LPBF machine. These sensitive regions lead to thermal accumulation due to the high density of input laser energy with great depth in adaptive scan patterns.

In our DRL framework, we have designed the reward function to directly guide the agent to avoid sensitive regions. The proposed DRL-based toolpath optimization algorithms do not involve the calculation of the depth of the molten pool. Instead, the reward functions utilize the turning angle to calculate the main term (introduced in Section 2.4), enhancing computational efficiency. The inputs for the data-intensive numerical model during the training process are the state (coordinate) of the agent, and the corresponding turning angles are calculated using geometric relations. Subsequently, rewards can be efficiently calculated using the equations introduced in Section 2.4. The codes for the data-intensive numerical model were written in the *Python* programming language, and for better implementation, we integrated these codes into our developed DRL framework. As a result, the time cost of the training process for 1000 episodes can be reduced from several hundred hours to around only one hour.

***Definition 1:*** A "sensitive region" in adaptive scan patterns is defined as two successive turning angles less than 90 degrees, with a length between them smaller than a particular value determined by the coefficient of the sensitive region. This coefficient is related to the fabrication environment and is significant because these regions result in thermal accumulation due to the high density of input laser energy with great depth in adaptive scan patterns.

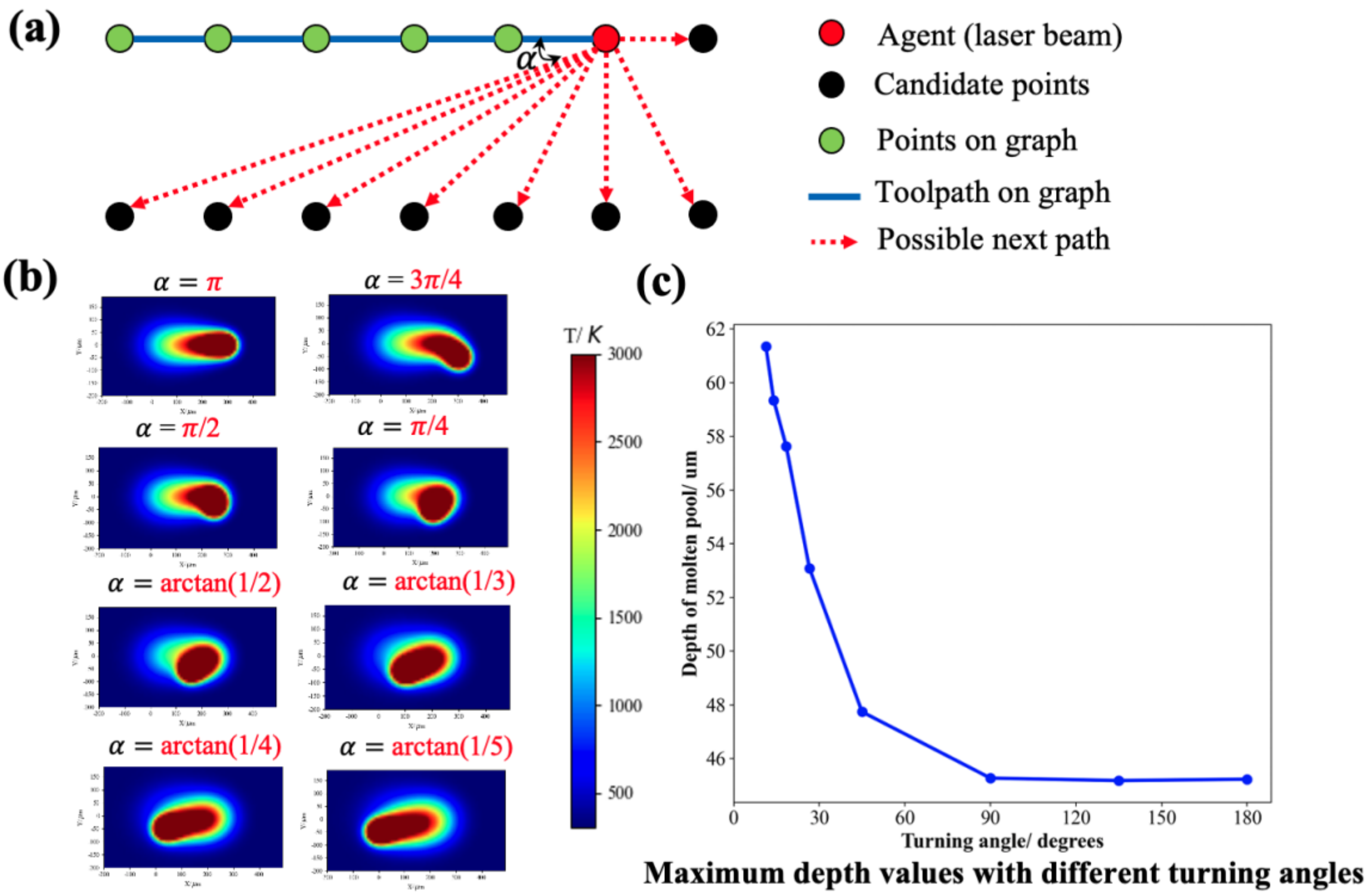

Fig. 2 (a) Toolpath with different possible turning angles (b) Temperature distribution results of 8 different turning angles (from $\pi$ to arctan $(1/5)$) (c) The relations between the maximum depth values of molten pool and different turning angles.

## 2.3 Action spaces design

To simplify the action selection, we set the searching radius to $\sqrt{2}h$, which enables us to search for the $k$ nearest neighbors (KNN) around the active points within the given searching radius. The agents are denoted by red colors, and the possible actions are depicted by red dashed lines with arrows, as shown in Fig. 3. There are three typical types of agents: free agent, edge agent, and corner agent. For instance, the free agent (Fig. 3 (a)) has seven possible choices: up, down, left, right, left-up, right-up, and right-down directions, excluding the direction along the previous path. If the agent is positioned at the corners of the printing domain, only two possible actions can be taken, which include up, down, left-up, and left-down directions (Fig. 3 (b)). If the agent is located at the boundaries, two possible actions can be taken, including up and left-up directions (Fig. 3 (c)), except for the left direction, which is the previous path's direction. It is important to note that these actions are theoretically feasible since the action selection should be performed inside the below-designed action spaces.

All points within the KNN set are used to calculate temperature values and angles between the previous path and all possible paths. The smoothing constraint described in [20] is applied to assess the smoothness of the candidate paths. Three actions have been designed in the action space, including selecting the path with the minimum temperature value, the smoothest path, or the second smoothest path. As introduced in the ATG algorithm [20], temperature values can be calculated using equation (4), which takes into account the time decay and the distance to the laser source. Equation (5) is used to choose the candidate path with the largest angle relative to the previous path (normalized between 0 and 180 degrees) to obtain the smoothest paths [20]. To avoid local optimal solutions when selecting the next path using equation (4) or (5), we can calculate the 2nd largest value using equation (6) as a relaxation factor. Therefore, the smoothest and second smoothest choices can be obtained using equations (5) and (6), respectively.

In Fig. 4 (a), all three possible actions are shown with dashed lines, including the right-up direction with minimized temperature values calculated by equation (4), the up direction with the smoothest path calculated by equation (5), and the left-up direction with the 2nd smoothest path calculated by equation (6). These three optimized selections can be used in the action space design problem as follows:

$$\boldsymbol{minimize} \quad F = \sum_{i=1}^{s} w_{ni}(t)\, T_i(d) \tag{4}$$

$$\boldsymbol{maximize} \quad |\alpha| \tag{5}$$

$$\mathbf{2^{nd}}\ \boldsymbol{largest} \quad |\alpha| \tag{6}$$

$$\boldsymbol{s.t.}\ 0 \le \alpha \le 180, 0 \le d \le \sqrt{(max\{x_i\})^2 + (max\{y_i\})^2}, t \ge 0$$

where $t$ is time, $d$ is distance, $T_i(d)$ is distance term, $w_{ni}(t)$ is normalized weight value related to time term, $s$ is the total number of points for calculation. $\alpha$ is the normalized angle between 0 to 180 degrees for the easy of comparison.

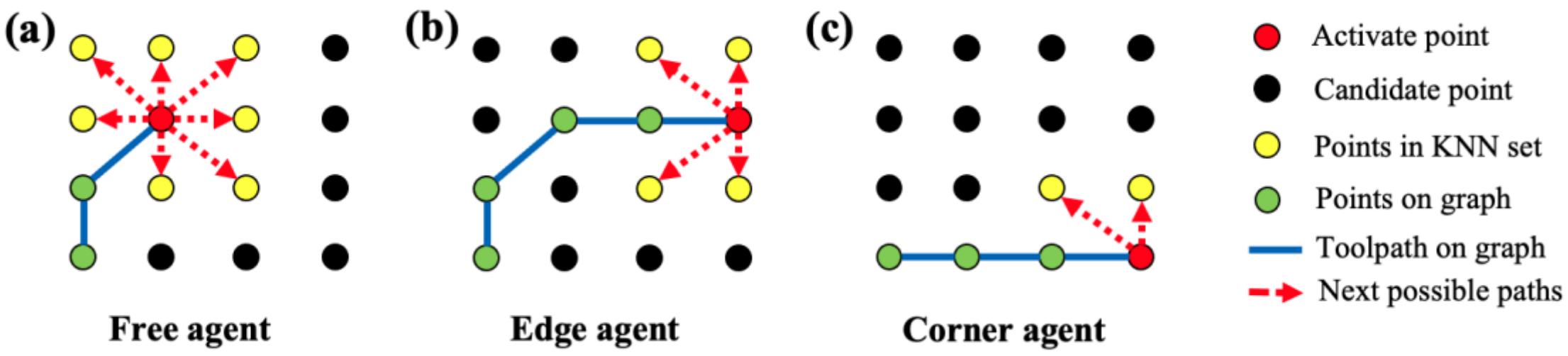

Fig. 3 Geometric constrain of agent movement (a) Free agent; (b) Edge agent; (c) Corner agent.

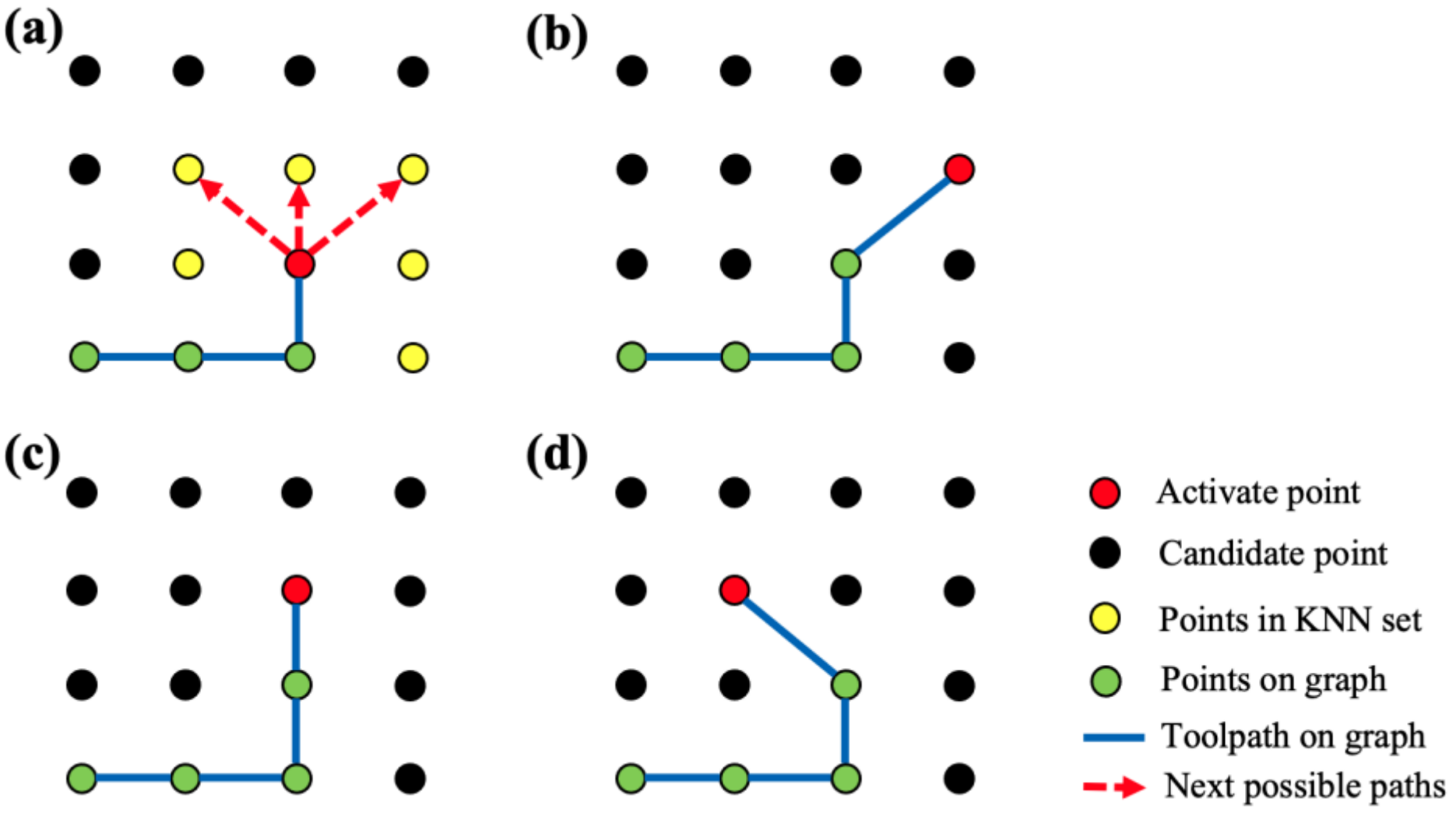

Fig. 4 Design of action spaces (a) Three possible paths (b) Next path with the minimum temperature (c) Next path with the smoothest angle (d) Next path with the 2nd smoothest angle.

### 2.4 Reward function design

Since the goal of the reward function is to reduce thermal accumulation and avoid deep overheating caused by accumulation on the temperature fields, the main term in the reward function (equation (7)) can be defined as the energy density over the affected regions, as shown in equation (8). To simplify the model and speed up the training process, we set the radius of affected region as $3h$ and simplify the affected area by considering the distance $d$, as shown in equation (9).

Apart from the main term, the reward function in our proposed algorithms also includes a regularization term. To avoid potential failure during printing, it is essential to minimize the occurrence of remelting caused by collisions in scan patterns. To achieve this, we perform "void-moving" (moving with laser off) to avoid collisions. However, excessive void-moving may result in an increase in fabrication time. Therefore, the regularization term takes collision times into consideration. The collision-related factor can be calculated using equation (10). During training, a judgment is made based on whether the current active point has been recorded as a collision more than a certain number of times. We count the collision times during the training process and compare it with a collision threshold, which we set to 3 in our experimental practice. If the current point exceeds the threshold, the reward value will be reduced by 1; otherwise, it will be 0.

Moreover, the reward function also considers the isolated-point factor, which will be introduced in Section 2.6. Therefore, the isolated-point also needs to be judged during the training process. If the current activated point has been regarded as the isolated-point, the reward value needs to be reduced by 1; otherwise, it will be 0. As a result, the reward function for each agent at time $t$ is shown in the following equations (7) – (11).

$$R = \sum_t (R_t - N_c - N_i) \tag{7}$$

$$R_t = -\frac{E_{input}}{A} \tag{8}$$

$$R_t = \begin{cases} -\frac{h}{d} & 0 \leq \alpha < 90 \; and \; d \leq 3h \\ 0 & otherwise \end{cases} \tag{9}$$

$$N_c = \begin{cases} 1 & n > 3 \\ 0 & otherwise \end{cases} \tag{10}$$

$$N_i = \begin{cases} 1 & isoloated - point \\ 0 & otherwise \end{cases} \tag{11}$$

where $R$ is the accumulated reward value; $R_t$ is the original reward value; $d$ is the distance between current sensitive angle and previous sensitive angle; $h$ is hatch distance of uniformly sampled points; $N_c$ is the collision penalty; $N_i$ is the isolated-point penalty; $n$ is the number of collision times; $E_{input}$ is the input laser energy; $A$ is the affected area of the input energy.

### 2.5 Exponential decay function

The $\varepsilon$-greedy strategy is commonly used in reinforcement learning algorithms, which involves selecting actions based on a comparison between a randomly generated value and $\varepsilon$, a value between 0 and 1 that represents the probability of choosing an action. If the randomly generated value is less than or equal to $\varepsilon$, the next action is chosen based on the Q value, otherwise, it is chosen randomly. The $\varepsilon$ value determines the balance between exploration and exploitation [31].

There are two extreme cases for $\varepsilon$: 1) if $\varepsilon$ is equal to 1, all actions are explored randomly; 2) if $\varepsilon$ is equal to 0, all actions are exploited based on the maximum Q value. To ensure a balance between exploration and exploitation, equation (12) is used to calculate the value of $\varepsilon$. The equation includes the current number of episodes $n$, the decay factor $\omega_e$, the starting epsilon value $\varepsilon_0$, the final epsilon value $\varepsilon_1$, and the current epsilon value $\varepsilon$. Therefore, the epsilon value is adjusted as the number of episodes increases, with the starting value decreasing exponentially towards the final value. This allows for a balance between exploration and exploitation, ensuring that reinforcement learning algorithms are not limited to learned experiences while also exploiting what has been learned.

$$\varepsilon = \varepsilon_1 + (\varepsilon_0 - \varepsilon_1)e^{-\frac{n}{\omega_e}} \tag{12}$$

where $n$ is current number of episodes; $\omega_e$ is the decay factor; $\varepsilon_0$ is starting epsilon value; $\varepsilon_1$ is final epsilon value; $\varepsilon$ is current epsilon value.

### 2.6 Other constraints

During the moving of agent inside the sampled points in our proposed algorithms, two more special environments should be considered, including unqualified-point and isolated-point.

***1) Unqualified-point:*** An active point is defined as an unqualified-point when its KNN points set is empty, indicating that no action can be taken within the set searching radius. For such points, the current KNN points set should be discarded, and a path generated by nearest distance algorithm should be found. Unqualified-points can be generated in two typical conditions: Fig. 5 (a) shows a typical example of an unqualified-point in red color located at the left-up corner restricted by previous toolpath and boundary, while in Fig. 5 (b), an unqualified-point could be totally restricted by the previously generated toolpath.

For the unqualified-point, the point in all the candidate points set with the nearest distance will be chosen as the next active point until all the points have been traversed. Since void-moving will be conducted when encountering unqualified points, the nearest neighbor candidate points are chosen to reduce the time costs during implementation. The distances of points in the candidate set to the activate point will be calculated, and the points with the minimum distance will be chosen, as shown in the following equation:

$$\begin{aligned} &\boldsymbol{minimize}\ \ f(d_i) \\ &\boldsymbol{s.t.}\ h \le d_i \ \le L \end{aligned} \tag{13}$$

where $d_i$ is the distance from the current activate point. $L$ is the maximum length of printing domain. $h$ is the hatch distance of uniformly sampled points.

***2) Isolated-point:*** An isolated-point can be defined as a point that requires two successive collisions to be connected - one to the previous path and one to the next path - or it is the last point connected by an unqualified-point. In Fig. 6, isolated-points are plotted in blue, unqualified points are plotted in red, and the path with collisions is plotted with red dashed lines. Two typical isolated-points are illustrated: isolated-points can be generated by the previous path and the next path resulting in two successive collisions in Fig. 6 (a), or by the last point connected by an unqualified-point in Fig. 6 (b). If isolated-points are melted, it can result in point-melting, leading to a significant increase in surface roughness. Consequently, void-moving is triggered when isolated-points are encountered, and the nearest neighbor candidate points are selected. However, this void-moving process does extend the printing time. Therefore, it is essential to minimize the number of isolated-points in scan patterns. To reduce the occurrence of isolated-points in scan patterns, we designed reward functions that include an additional penalty term (introduced in Section 2.4).

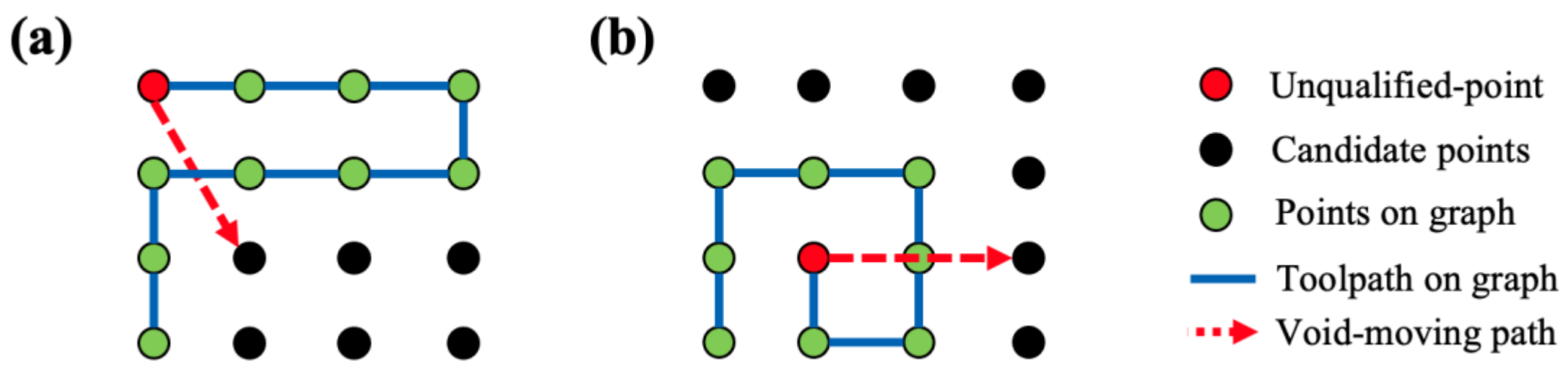

Fig. 5 Examples of unqualified-points and void-moving (a) Unqualified-point restricted by the previous toolpath and boundary (b) Unqualified-point restricted by the previous toolpaths.

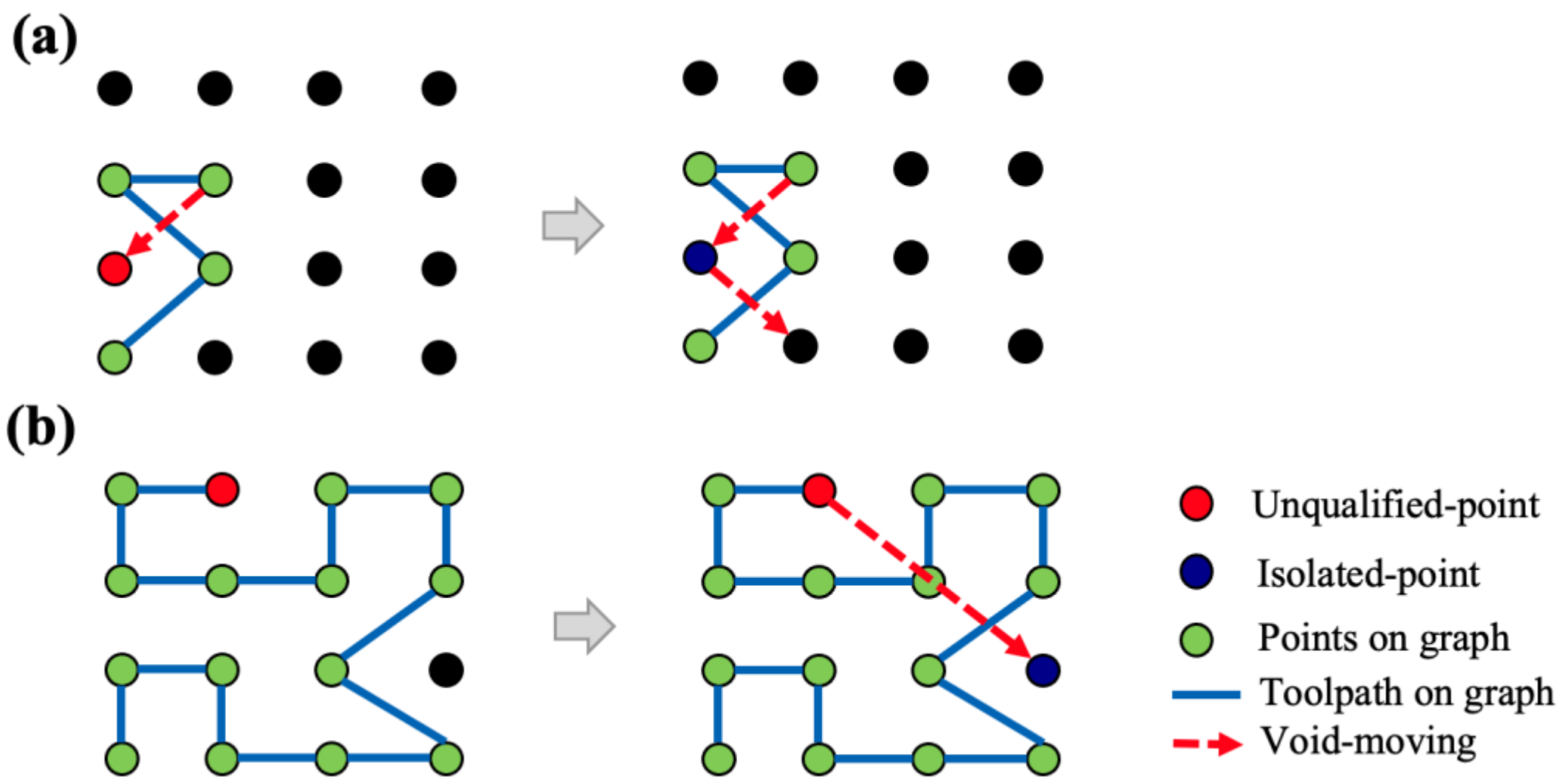

Fig. 6 Examples of isolated-points (a) Isolated-point generated by two successive collisions (b) Isolated-point generated by an unqualified-point.

## 3. Algorithm implementation

The neural network architectures and input-output relationships are illustrated in Fig. 7. The states, represented as coordinates, serve as the input to the neural networks. Both the main and target models consist of three fully connected layers, including two hidden layers with 24 and 12 units, respectively. The two models are initialized identically, and the target model is updated by the main model at a predetermined frequency, provided that the stored data exceeds the replay size. The Rectified Linear Unit (ReLU), a commonly used activation function in deep learning models, is employed as the activation function for the hidden layers. The movement of the laser agent is guided by the maximum Q values from the output layer.

Table 1 presents the main parameters and their corresponding values in the neural network architectures. These parameters were optimized through trial and error. The code is implemented in the *Python* programming language, using the *TensorFlow* framework for designing the neural network architectures. Throughout the training process, toolpath patterns were saved every 20 episodes, and the reward function curves were plotted. Eventually, optimal policies for generating the optimized scan patterns in these printing domains were successfully obtained.

Fig. 8 shows an example of the implementation of DRL based algorithm on a rectangular printing domain. Fig. 8 (a) shows the training process in terms of rewards and the number of sensitive regions. The results demonstrate that both the reward and the number of sensitive regions converge to 0 after approximately 800 episodes, indicating that the final scan pattern is stable and does not include sensitive regions resulting in accumulated thermal fields. The selected typical episodes are depicted in Fig. 8 (b), ranging from 100 to 1000. It can be observed that the scan pattern can be obtained after completing 1000 episodes without any sensitive regions. The scan patterns during the first 500 episodes are somewhat disorganized, and there are a few sensitive regions, which can be calculated by the reward function as negative values.

The final generated scan pattern is identical to that of the ATG algorithm, which has been demonstrated to have advantages over commonly used zigzag scan patterns [18]. This case demonstrates that for simple cases, the DRL-based toolpath generation algorithm can obtain the same patterns as the ATG algorithm. To further illustrate the advantages of the DRL-based algorithm compared to the ATG algorithm, numerical simulations with typical polygon shape will be conducted in Section 4 and experiments will be conducted in Section 5.

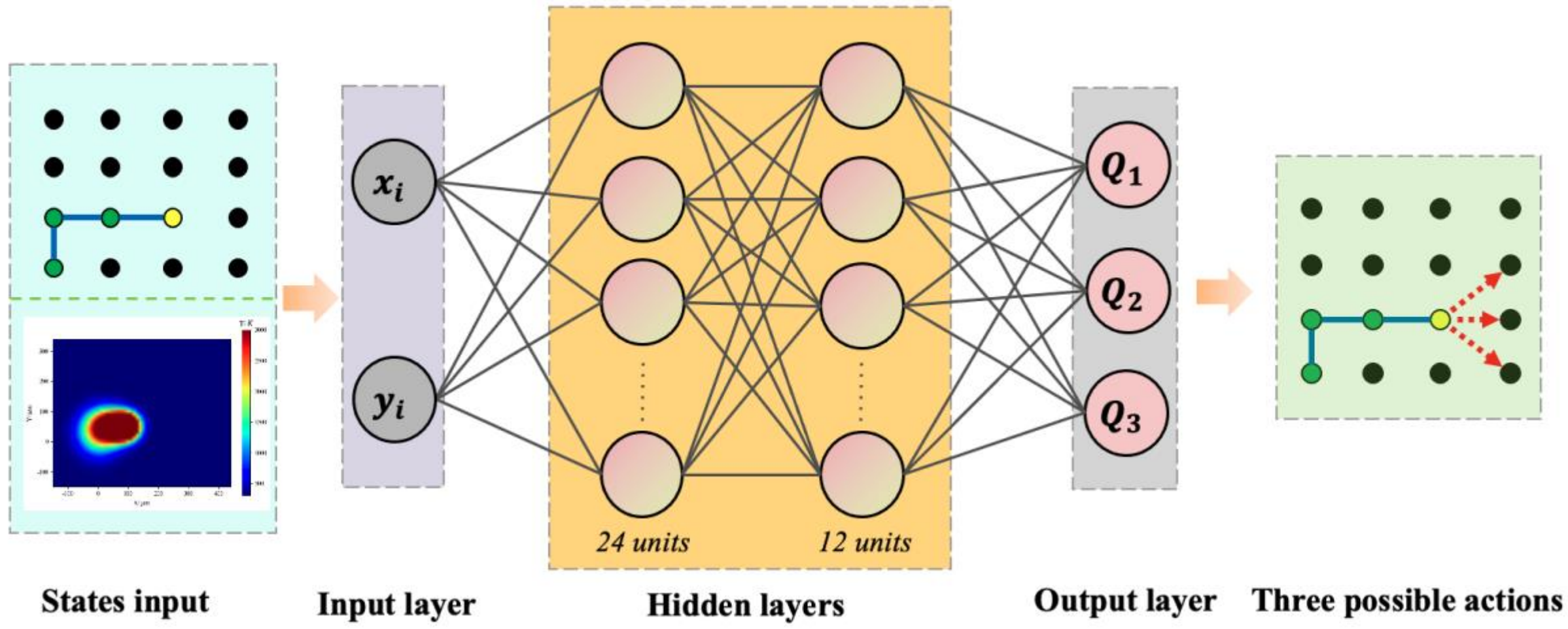

Fig. 7 Neural network architectures.

Table 1 Main parameters in the neural network architectures

| Parameters | values |
|---|---|
| Learning rate $\alpha$ | 0.001 |
| Discount factor $\gamma$ | 0.99 |
| Batch size | 64 |
| Update frequency | 80 |
| Replay size | 1000 |
| Initial epsilon $\varepsilon_0$ | 1 |
| Final epsilon $\varepsilon_1$ | 0 |
| Decay coefficient $\omega_e$ | 200 |

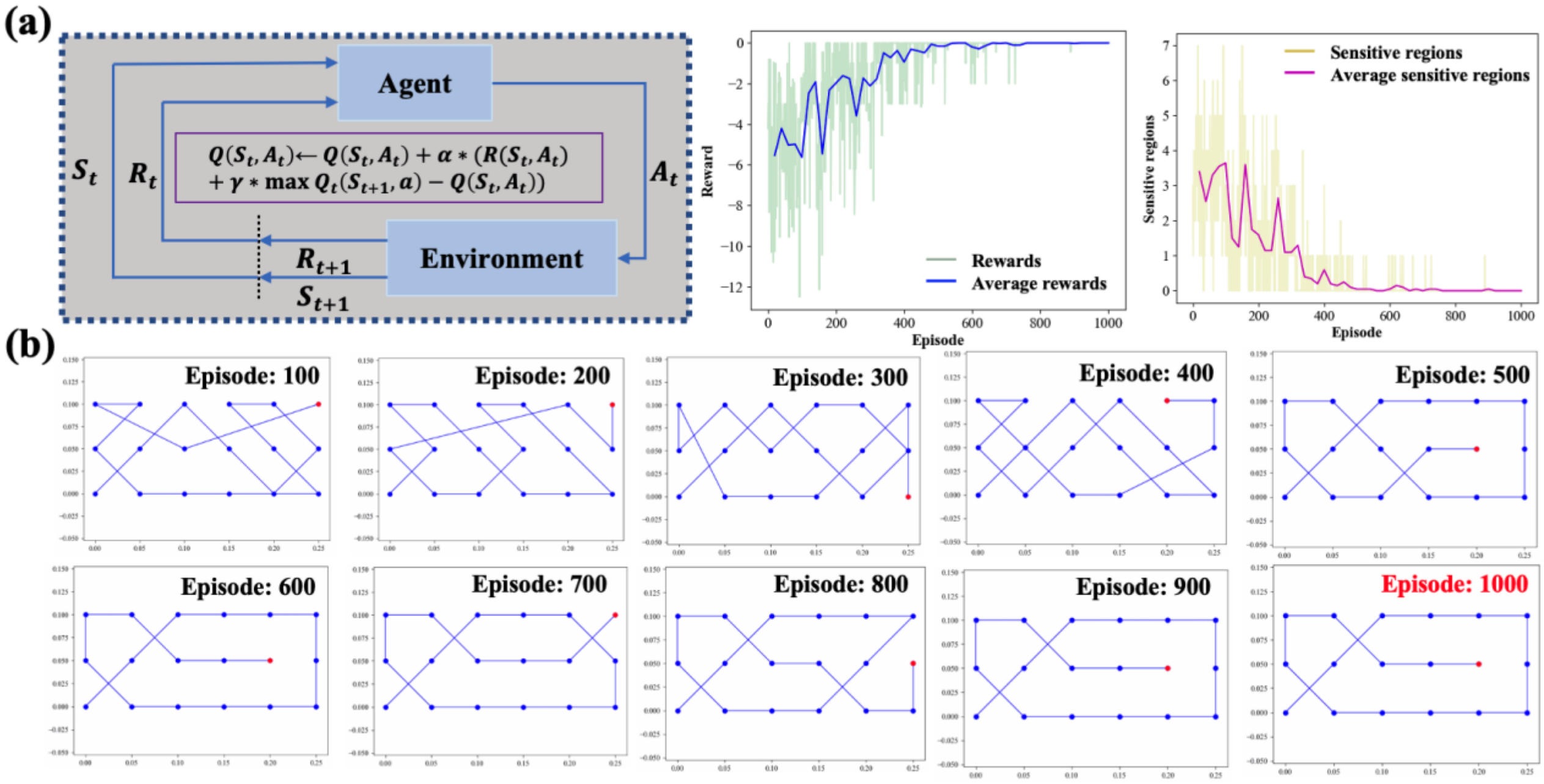

Fig. 8 Implementation of DRL based algorithm (a) Iteration of DRL based algorithm and reward and sensitive region curves (b) Scan patterns generated during the training process.

## 4. Numerical simulation for polygon shape case study

To further demonstrate the advantages of our approach, we conducted the general case study with polygon shape in this section, which is commonly encountered in fabrication processes. We utilized a DRL-based algorithm to generate scan patterns that avoid extreme thermal accumulation regions. The generation process, including the initial sampling points, several scan patterns generated during the training process, optimized scan patterns, and reward curves, are presented. Furthermore, we included scan patterns generated by the zigzag method and the ATG algorithm for comparison. We conducted numerical simulations of these three strategies and plotted the depth values of the molten pool during the simulated fabrication process. We compared the average and peak values obtained using each strategy as well.

Fig. 9 shows the numerical simulation of three different toolpath patterns for the polygon shape printing domain. The process of generating the optimized scan pattern is depicted in Fig. 9 (a) using 1000 episodes, from the initial sampled points to the final optimized scan pattern. The reward during the training process converged to 0 as the number of episodes increased. The optimized scan pattern obtained at 600 episodes does not contain any sensitive regions. In Fig. 9 (b), the optimized scan pattern generated by the DRL algorithm is compared with the scan patterns generated by the ATG algorithm and Zigzag patterns, which start from the same position at the left-down corner. The Zigzag patterns are represented by black lines and start along the X direction, while the scan patterns generated by the ATG algorithm are shown in blue lines in the middle of Fig. 9 (b). The scan pattern generated by the DRL algorithm is plotted in red lines on the right of Fig. 9 (b). Numerical simulations using the RUSLS method are performed to calculate the temperature fields during the movement of the laser beam along the scan patterns, and the depth values of the molten pool are recorded for comparison.

Fig. 9 (c) presents a comparison of the depth of the molten pool for all three scan patterns. The DRL-based algorithm is represented in red, the ATG algorithm in blue, and the zigzag pattern in black. The average values are depicted with dashed lines, and the peak values are indicated by inverse triangle marks of the corresponding colors. From the results, we can conclude that although the ATG algorithm reduces the average depth values by approximate 6% compared to the zigzag scan pattern, the DRL algorithm further reduces it by about 13%. Moreover, the peak depth values achieved by the DRL algorithm (78.6 $\mu m$) are smaller than those of the ATG algorithm (82.0 $\mu m$) and the zigzag pattern (83.3 $\mu m$). Additionally, the convergence of final reward values to 0 indicates the absence of sensitive regions in the DRL optimized scan patterns. These results indicate that our proposed DRL algorithm achieves an overall uniform thermal distribution while effectively avoiding extreme thermal accumulations. In contrast, the Zigzag scan patterns do not consider the thermal distribution, despite their convenience and simplicity. The scan pattern generated by the ATG algorithm may result in suboptimal outcomes due to the presence of accumulated thermal regions.

Therefore, we can conclude that our proposed DRL-based algorithms successfully generate optimized scan patterns that avoid extreme thermal accumulation regions in commonly used printing domains. By comparing three typical scan patterns, including the commonly used zigzag pattern and the advanced ATG algorithm, we illustrate the process and advantages of our approach. The results demonstrate that the optimized toolpath exhibits the most uniformly distributed temperature fields and minimizes the occurrence of accumulated temperature fields, effectively addressing the issues encountered with scan patterns generated by zigzag or ATG algorithms. The achievement of uniformly distributed temperature fields is a significant outcome of our work.

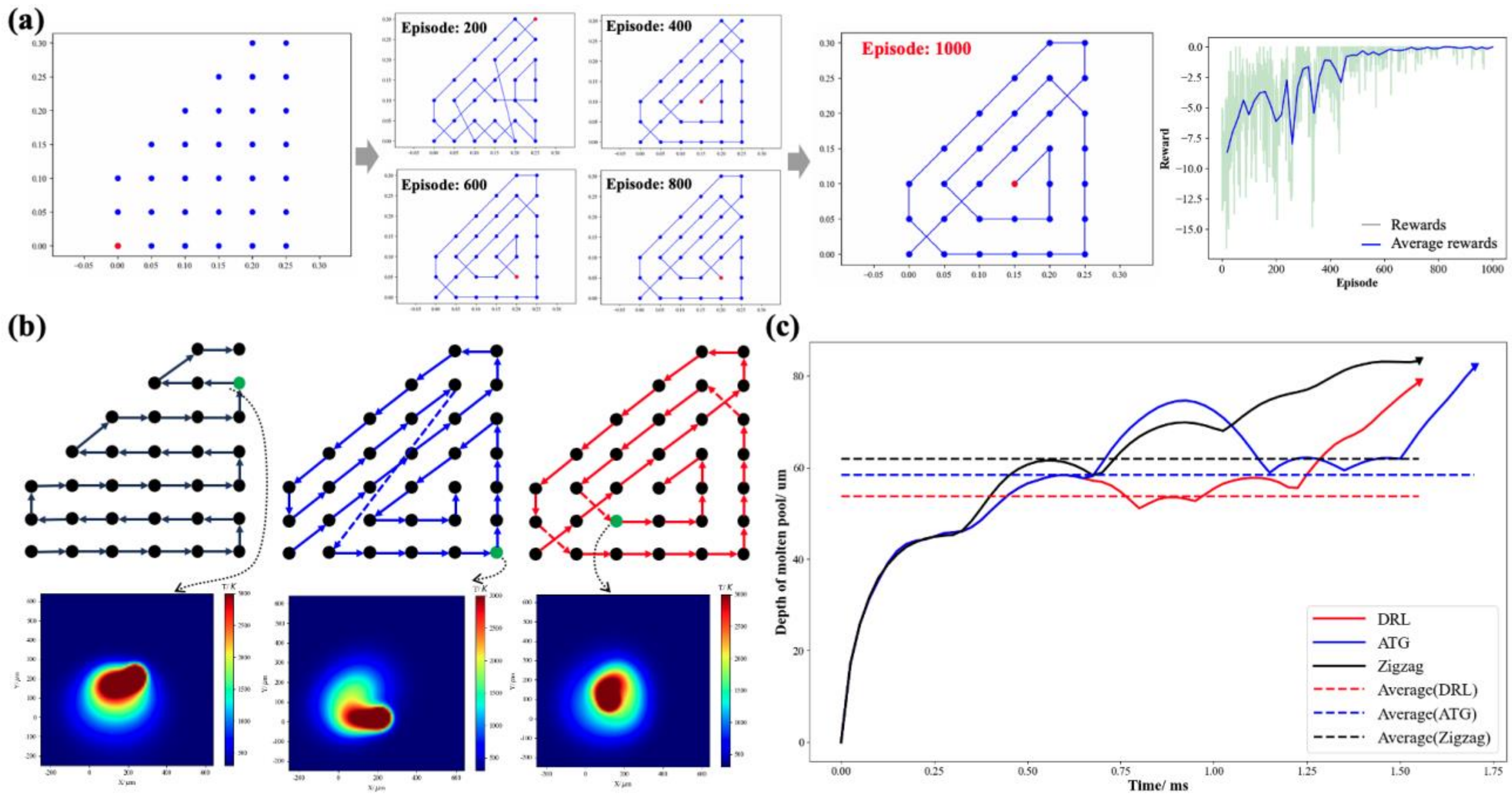

Fig. 9 Numerical simulation of three different toolpath patterns for the polygon shape printing domain (a) Scan pattern generation process of the DRL-based algorithm and reward curve (b) Comparison of the zigzag, ATG algorithm, and DRL algorithm in terms of scan patterns and their temperature fields in numerical simulations (c) Comparison of the depth of molten pools during the simulated fabrication process, with average values shown as dashed lines and peak values indicated by inverse triangle marks.

## 5. Experiments and results

### 5.1 Sample design

To prove the effectiveness of the proposed approach, we design the scan patterns within randomly generated Voronoi diagrams with polygon boundaries and then compare them with commonly used and advanced scan patterns (ATG algorithm). To accelerate the scan patterns generation process, we adopt the island-based strategy proposed in [20]. This approach involves dividing the printing domain into several islands and generating a sequence for these islands. In this case, the printing domain with a size of $30mm \times 30mm$ is separated into 36 islands with island element size at $5mm \times 5mm$.

The Voronoi diagrams are generated inside each island. To minimize the effects of randomness, three different Voronoi diagram cases are adopted and plotted in Fig. 10 (a). Each Voronoi diagram consists of 14 seeds in total, including 4 rectangular vertices and 10 randomly generated points inside the island. Then 14 polygons will be generated within each island, with the marking number from B1 to B14, as shown in Fig. 10 (a). Then the uniform sampling is conducted within each polygon. The trained model for polygon shape is adopted to generate DRL patterns. For example, DRL patterns inside of polygon B14 are shown in Fig. 10 (a). In addition, the ATG patterns are generated within the Voronoi diagrams as well by utilizing the algorithms introduced in [20]. After all the polygons have been traversed, the scan patterns within one island element can be obtained by merging the toolpath patterns. Finally, the sequence of all the islands of both DRL patterns and ATG patterns is planned by the sequence planning algorithm for island-based strategy [20]. Finally, the toolpath patterns should be converted to G codes, which will be recognized by the LPBF machine. We fine-tune the G codes by removing isolated points to avoid “point-melting”, which greatly affects the surface quality. During void-moving, the laser power is turned off before a collision occurs and turned on after moving to the next point. The pseudocodes are given in Algorithm 2.

Fig. 10 (b) shows the design of the fixtures for the LPBF machine. The thin plates in SS316L material with a thickness of 200 $\mu m$ were fixed on the substrate. Two commonly used scan patterns, including Zigzag patterns and Chessboard patterns [32], are used for comparison, as shown in Fig. 10 (c). Both Zigzag and Chessboard patterns are directly generated by the commercial software *Materialise Magics (Materialise NV, Lovaine, Belgium)*. The Zigzag patterns were created without islands, consisting only of long scanning lines parallel to each other. Chessboard patterns were generated with islands filled with short scanning lines. The detailed sequence for Zigzag patterns, Chessboards patterns, ATG & DRL patterns are given in Fig. 11. The line sequence for Zigzag patterns and islands sequence for Chessboard patterns are shown in Fig. 11 (a) and Fig. 11 (b), respectively. These sequences are generated directly by the software *Materialise Magics*. For ATG and DRL patterns, their islands sequences are calculated using the island-based strategy introduced in [20], as illustrated in Fig. 11 (c). A self-developed micro LPBF machine (Han’s Laser M100μ) is adopted in our research. Parameters are given in Table 2 for the LPBF machine. Four groups of scan patterns in Fig. 10 (d) are implemented on the thin plates.

**Algorithm 2: Toolpath generation within Voronoi diagrams**

**Input:** Number of vertices $n$, trained model $M$, threshold $t = \sqrt{2}h$
**Output:** G-codes for LPBF machine

1. Define the size of the island element
2. Randomly generated Voronoi diagrams within the islands
3. ***for*** each polygon ***in*** Voronoi diagrams:
4.     Uniformly sample points inside the polygons
5.     Generate toolpath patterns based on the trained model $M$
6.     Calculate the average coordinates $(\overline{x}, \overline{y})$
7. Merge the Voronoi polygons inside the islands
8. Plan the sequence of islands using inverse distance weighting approach
9. Convert the toolpath patterns into G-codes
10. ***for*** each coordinate $(x_i, y_i)$ ***in*** the G-codes # Fine-tune the G-codes
11.     Calculate the previous and next distance $d_{i-1}$, $d_{i+1}$
12.     ***if*** abs($d_{i+1}$) > $t$: off the laser power after $(x_i, y_i)$
13.     ***if*** abs($d_{i-1}$) > $t$ ***and*** abs($d_{i+1}$) > $t$: remove $(x_i, y_i)$
14. ***Return*** the G-codes for LPBF machine

Table 2 Parameters of LPBF machine

| Parameters | Value |
|---|---|
| Laser power (W) | 50 |
| laser diameter ($\mu m$) | 25 |
| Scanning velocity ($mm$/s) | 1000 |

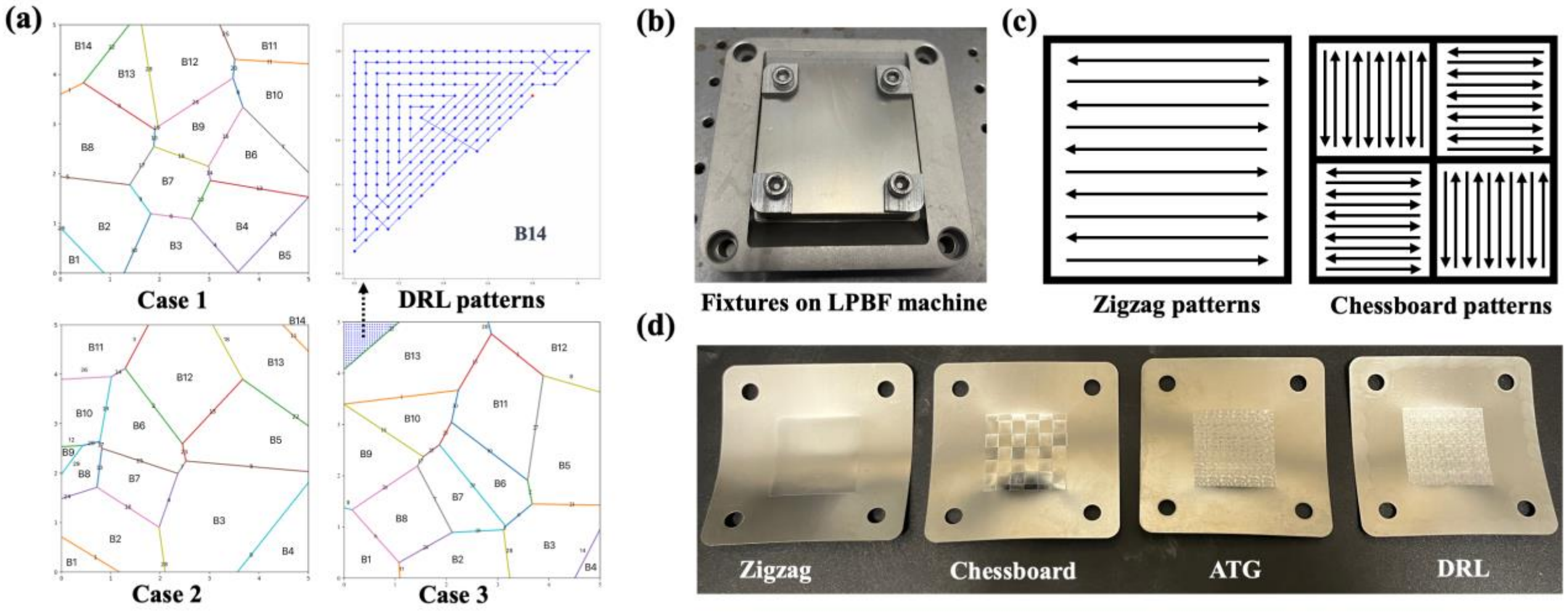

Fig. 10 Samples with different scan patterns implemented by the LPBF machine (a) Three randomly generated Voronoi diagrams and DRL patterns generated inside polygons (DRL patterns inside B14 of case 3 are shown as an example) (b) Fixtures on the LPBF machine (c) Designed Zigzag and Chessboard patterns for comparison (d) Four groups of thin plate samples with different toolpath patterns.

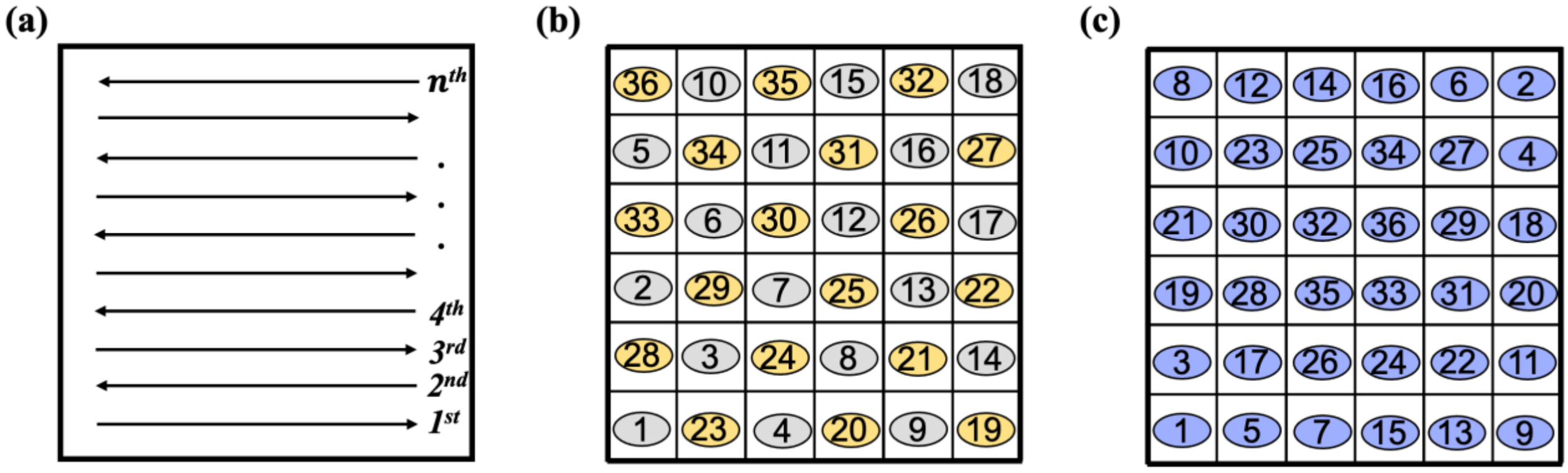

Fig. 11 (a) Scanning sequence for Zigzag patterns; (b) Island sequence for Chessboard patterns (grey and yellow colors represent two scanning lines in perpendicular directions); (c) Islands sequence for ATG and DRL patterns calculated by island-based strategy introduced in [20]

### 5.2 Results

To measure the maximum distortions, four groups of samples were compared by fixing the side of thin plates. Both the front view, left-side view and right-side view are depicted in Fig. 12. The maximum distortions are indicated using green, yellow, blue and red arrows, representing the samples with Zigzag patterns, Chessboard patterns, ATG patterns and DRL patterns, respectively. The Zigzag patterns exhibit the largest maximum distortion at 18 $mm$, the Chessboard patterns show the second largest maximum distortion at 13.5 $mm$, the maximum distortion of ATG patterns is 11.5 $\pm$ 0.5 $mm$, and the DRL-optimized patterns exhibit the smallest maximum distortion at 9.5 $\pm$ 0.7 $mm$. It can be concluded that the plate with the toolpath generated by our proposed DRL algorithms (indicated by the red arrow) exhibits a reduction in maximum distortion of approximate 47% when compared to Zigzag patterns (indicated by the green arrow), approximate 29% when compared to Chessboard patterns (indicated by the yellow arrow), and approximate 17% when compared to ATG patterns (indicated by the blue arrow).

The deformation of thin plate samples is directly affected by different scan patterns. In the case of the sample with Zigzag patterns, it exhibited the largest distortion due to the unevenly distributed thermal fields during the movement of laser beam, resulting in an accumulation of residual stress. Chessboard patterns with islands can help mitigate the issue of long scanning lines when compared to Zigzag patterns but still exhibit the second-largest distortion. Compared with both Zigzag and Chessboard patterns, scan patterns generated by the ATG algorithm can obtain smaller distortions by minimizing the thermal gradients. However, there may be still some extremely accumulated temperature fields, as shown in numerical simulation analysis in the Section 4. In contrast, our proposed DRL-based scan patterns generation approach can achieve significantly less distortion because they generate a more uniformly distributed temperature field and minimize the accumulation of residual stress. These experimental results provide compelling evidence for the effectiveness of our proposed DRL-based toolpath generation framework.

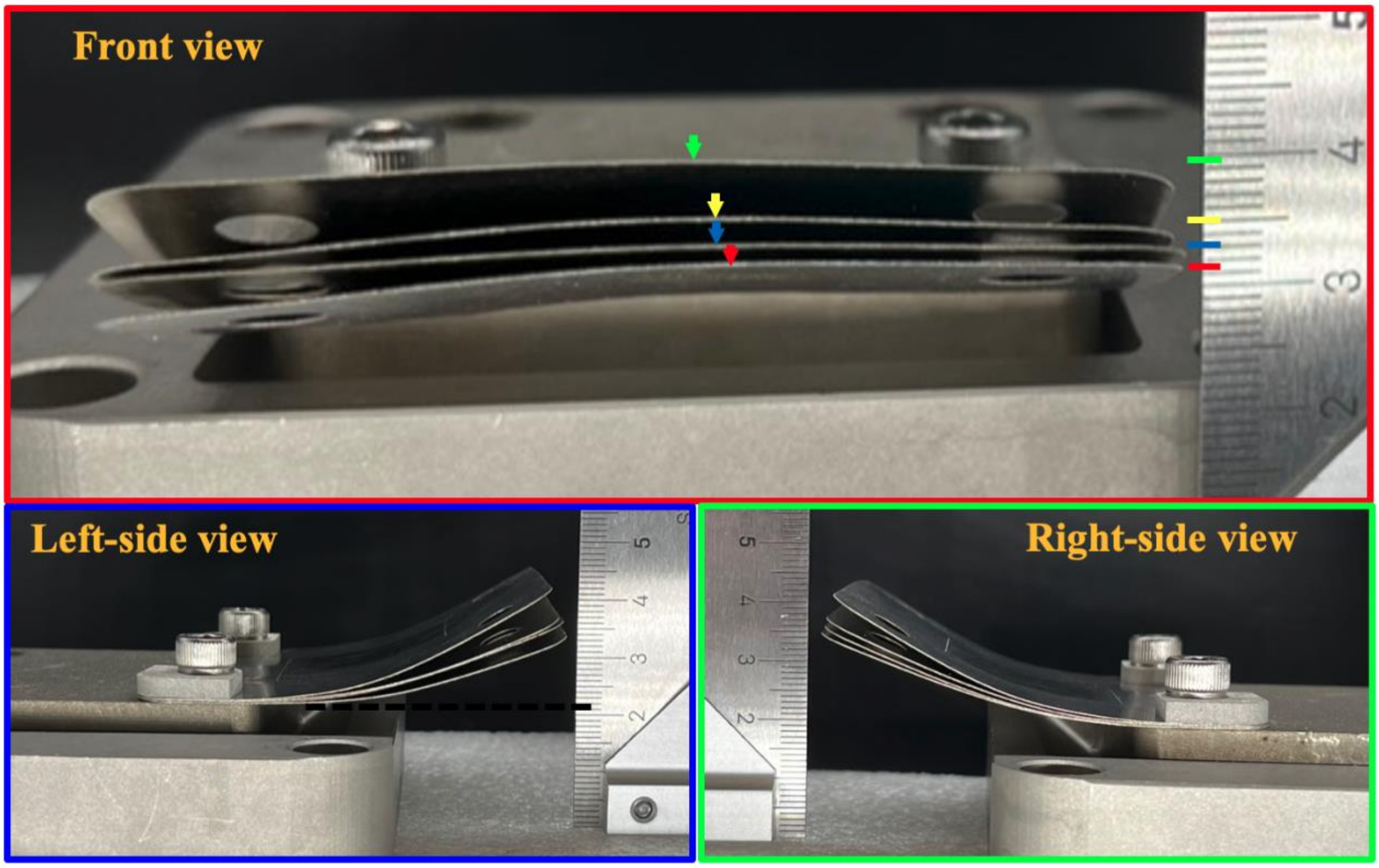

Fig. 12 Comparison of the distortions between four groups of samples in front view, left-side view and right-side view (The maximum distortion of Zigzag patterns, Chessboard patterns, ATG patterns and DRL patterns are shown in green, yellow, blue and red arrows, respectively).

## 6. Conclusions and future works

The paper presented a DRL-based toolpath generation framework to generate scan patterns with the most uniformly distributed temperature fields. A simplified numerical model was developed by considering the relationship between turning angles and thermal distributions to improve computational efficiency. A reward function was designed with the aim of minimizing the input energy density, and the parameters in the neural network architecture were determined through experimental trial-and-error approaches. Special environments for the agent, such as unqualified-point, isolated-point, and sensitive regions, were defined during the training of scan patterns. The paper included numerical simulation for the polygon case study that demonstrated the effectiveness of the DRL-based adaptive toolpath optimization framework in obtaining uniformly distributed temperature fields and avoiding excessive thermal accumulation. The experimental results also proved that our proposed algorithm could greatly reduce deformation compared with Zigzag patterns, Chessboard patterns and ATG patterns. In summary, the study presented a promising approach for using machine learning to optimize the toolpath patterns in the LPBF process and provided a foundation for further research in this area. The following main conclusions can be drawn from the study:

1) A framework based on the DRL algorithm was developed to generate optimized scan patterns in the LPBF process. The objective of the framework was to minimize the energy density and achieve uniformly distributed thermal fields.
2) Numerical simulation results for the polygon shape case demonstrated that the DRL algorithm could reduce thermal accumulation by approximate 13% compared with the zigzag scan pattern. In addition, the occurrences of accumulated temperature fields can be reduced to 0 compared with the ATG algorithm.
3) Experimental results showed that the plate with the toolpath generated by our proposed DRL algorithms exhibits a reduction in maximum distortion of approximate 47% when compared to Zigzag patterns, approximate 29% when compared to Chessboard patterns, and approximate 17% when compared to ATG patterns.

4) Both the numerical simulation and experimental results proved the effectiveness of utilizing machine learning algorithms in the toolpath generation for LPBF processes.

There are still areas for future work. Firstly, the consideration of inter-layer thermal accumulation in 3D geometries can be explored by designing new reward functions. The impact of noise points generated around the boundary due to precision issue in uniform sampling needs to be addressed to ensure the robustness of the trained model and further reduce the number of isolated-points. Currently, all paths are composed of straight line segments. We will investigate the use of curved segments to enhance our optimization algorithms, which may also help eliminate noise points.

Due to limitations in computational resources, the model was trained on regular polygon shapes. It will be enhanced by adding more training data from complex environments, including those with free-form boundaries. Moreover, further research should be conducted on shape and number optimization of polygons in Voronoi diagrams. Computational efficiency will be enhanced to accelerate the toolpath generation process, particularly for more complex structures. Considering that the argon flow, smoke, or spatter generated from the LPBF process may affect the selection of optimal candidate points, future algorithm improvement will take these factors into consideration. In addition, it's worth noting that the adoption of the void-moving operation for unqualified points may increase fabrication time. The presence of noise points around the boundaries of randomly generated Voronoi diagrams also intensifies the occurrence of void-moving. Therefore, the efficiency of fabricating large parts should be studied in the future.

## CRediT authorship contribution statement

**Mian Qin:** Conceptualization, Methodology, Formal analysis, Investigation, Software, Validation, Visualization, Writing – original draft, Writing – review & editing. **Junhao Ding:** Methodology, Writing – review & editing. **Shuo Qu:** Methodology, Writing – review & editing. **Xu Song:** Conceptualization, Methodology, Investigation, Resources, Writing – review & editing. **Charlie C.L. Wang:** Conceptualization, Investigation, Writing – review & editing. **Wei-Hsin Liao:** Supervision, Conceptualization, Methodology, Investigation, Resources, Funding acquisition, Writing – review & editing.

## Acknowledgments

This work is supported by the Research Grants Council (Project No. C4074-22G) of Hong Kong Special Administrative Region, China, and The Chinese University of Hong Kong (Project ID: 3110174).